\documentstyle[prl,aps,12pt]{revtex}
\begin{document}
\draft
\title{\Large\bf  Definition of Chern-Simons Terms in Thermal $QED_3$ Revisited}
\author{S. Deser$^{(a)}$,   L. Griguolo$^{(b)}$  and 
        D. Seminara$^{(a)}$\footnote{
{\tt Deser, Seminara@binah.cc.brandeis.edu; Griguolo@irene.mit.edu}}}
\address{\it $^{(a)}$ Department of  Physics, Brandeis University,
                        Waltham, MA 02254, USA\\
             $^{(b)}$ Center for Theoretical Physics, Laboratory
                         for  Nuclear Science and Department of Physics,\\
                         Massachusetts Institute of Technology, Cambridge,
                         Massachusetts 02139, U.S.A.}
\date{Received \today}
\maketitle
\medskip
\font\ninerm = cmr9
\pagestyle{empty}
\begin{abstract}
\ninerm\noindent
We present two compact derivations of the correct definition of the 
Chern-Simons term in the topologically non trivial context of 
thermal $QED_3$. One is based on  a transgression descent from a $D=4$ 
background connection, the other on embedding the abelian model in $SU(2)$. 
The results agree with earlier cohomology conclusions and can be also used 
to justify a recent simple heuristic approach. The correction to the naive 
Chern-Simons term, and its behavior under large gauge transformations are 
displayed.
\end{abstract}
\pacs{PACS numbers:\ \  11.10.Wx 11.15 11.30.Er 11.30RD\hfill}
\vfill
\begin{flushright}
BRX-TH-425
\end{flushright}
\newpage
\pagestyle{plain}
Chern-Simons (CS) terms, defined in odd dimension, contain gauge 
information not accessible through the field strength alone. We have in 
mind large gauge transformations, ones not shrinkable to  the identity, 
to which CS (but not $F_{\mu\nu}$) is sensitive. However it is also known 
\cite{Poly,DW,US,Orlando} that, as normally expressed, the abelian CS 
term (in $D=3$, say)
\begin{equation}
\label{ICS}
I_{CS}=\frac{1}{8\pi^2}\int A\wedge F
\end{equation}
is not always well-defined, but requires corrections
\footnote{The non-abelian case differs in a number of respects; it 
will addressed separately \cite{us2}.}. Our twofold aim is to 
obtain the correct form, in two complementary, compact, ways and to show 
explicitly that ``improvement'' of $I_{CS}$ is already needed in simple but 
quite physical contexts, such as  abelian $U(1)$ gauge fields in $D=3$ with 
non-trivial topology. A
generic example is finite temperature $QED_3$, where $t$ ranges over a finite 
circle $S^1$ of perimeter $\beta=1/\kappa T$ and space is a closed 
$2-$manifold $\Sigma^2$ with associated non-vanishing magnetic flux $\Phi=
\int dx^2 B,$ $B\equiv\displaystyle{\frac{1}{2}\epsilon^{ij} F_{ij}=F_{12}}$.
[We recall that the flux is a necessarily quantized topological
invariant, $\Phi=2\pi k$; see {\it e.g.} \cite{math}.] 
The need to improve the naive $I_{CS}$ is due to the fact that  it 
explicitly involves  the vector potential $\bf A$, ``modulated'' by the field 
strength. But presence of magnetic flux implies that $\bf A$ will depend on 
the patches needed  to cover the closed manifold $\Sigma^2$; hence the 
integral in 
(\ref{ICS}) as it stands will be, unacceptably, patch-dependent. This 
difficulty has been recognized and cured long ago both in cohomological 
$D=3$ calculations \cite{Poly,Orlando} and by descent from $D=4$
\cite{DW}; recently we have given a heuristic approach to the solution
\cite{US}.  Improvement of $I_{CS}$ is not a merely a mathematical nicety,
but  has  direct bearing on real $QED_3$ questions such as  the necessity and 
amount of quantization of its coefficient when $I_{CS}$ is viewed as a 
dynamical field action. Our present interest in 
$I_{CS}$ was aroused by calculations of effective 
QED actions induced by charged fermions, and the complex of questions raised 
there about the seeming appearance of induced CS terms and their coefficients
\cite{US}. Here we will first present a different
route to the (same) correct definition of $I_{CS}$, based on the Chern-Weil
theorem using the transgression formula involving
a background  connection $\hat A_\mu$ on the non-trivial bundle,  that 
compactly replaces the 
patch-dependence and the associated  boundary ``counter-terms'', by a simpler 
$\hat A_\mu-$dependent addition. We will then compare this method with the two
earlier approaches and use it also to justify  
the simple-minded ``derivation''. Finally, we give what is perhaps a still 
easier definition by use of nonabelian embedding to take advantage of the 
simpler (!) cohomological properties avaible there.

We  begin our analysis from the usual
$4-$dimensional identity that leads to the introduction of the
abelian CS form,
\begin{equation}
\label{FF}
F\wedge F\equiv d(A\wedge F) \; .
\end{equation}
One cannot apply the Poincar\'e lemma to this identity when $A_\mu$ is 
nontrivial, as when it carries a nontrivial magnetic flux through 
$\Sigma^2$.  For then $F\wedge F$, while closed, is not exact; equivalently 
$A_\mu$ is globally defined on ${\cal M}_4$ as a connection, but not as 
a $1-$form. We circumvent this obstacle through the Chern--Weil theorem
(see {\it e.g.}, \cite{math}) which states (for our case) that, if ($F$, $\hat F$) are field 
strengths corresponding to two different connections ($A$, $\hat A$) on some
bundle, then ($F\wedge F-\hat F\wedge \hat F$) is exact as well. A corollary,
the transgression formula, provides the explicit  $3-$form whose divergence
it is:
\begin{equation}
F\wedge F-\hat F\wedge \hat F= d\left[(A-\hat A)\wedge(F+\hat F)\right],
\end{equation} 
as is easily verified since the cross-terms on the r.h.s. cancel. We can
therefore 
define $I_{CS}$, also on non-trivial bundles, to be\footnote{In general a 
$D=3$ bundle is not a boundary of a $4-$bundle but cochains may be required
\cite{DW}. This complication does not occur in our explicit examples, but
can be handled as well.}
\begin{equation}
\label{ICS2}
\bar I_{CS}\equiv\frac{1}{8\pi^2}\int_{{\cal M}_4} 
\hat F\wedge \hat F+\frac{1}{8\pi^2} \int_{\partial {\cal M}_4} 
(A-\hat A)\wedge(F+\hat F)\; .
\end{equation}
The
explicit dependence of $\bar I_{CS}$ on $A-\hat A$ insures that it is
globally well defined: recall that a bundle is defined by the gauge transition
functions between patches, all connections
on the bundle having the same patch behavior. In particular, $A$ and $\hat A$ carry the same flux through
$\Sigma^2$; see also discussion after (\ref{coppola}).

We digress for a moment to note that appearance of an intrinsic reference
background is common in connection with (gauge or gravitational) anomalies
in non trivial topologies. What makes $\hat A$ unusual is that while it
transforms as a connection when changing patches -- so as to neutralize
the same behavior in $A$ --  we may (and do) choose it not to transform
under gauge tranformations that affect $A$ only; this too is not unknown,
for example in background field expansion of QFT. These different roles
for $A$ and $\hat A$ can be justified in terms of the usual BRST analysis
(see {\it e.g.}, \cite{anomalies}).

Returning now to $\bar I_{CS}$, the elegant aspect of (\ref{ICS2}) is its
``covariance'' (no patch dependence), paid for by its apparent dependence
on the $D=4$ background $\hat A$. In the other approaches there is no $\hat A$,
but covariance is lost. One gratifying property of (\ref{ICS2}) is that it
immediately reproduces the correct gauge variation of $\bar{I}_{CS}$ at finite 
temperature. Under the  large gauge transformation $A_0\to A_0+2\pi n/\beta$, 
${\bf A}\to {\bf A}$, $\bar I_{CS}$ changes  proportionally to the flux,   
\begin{equation}
\label{pizza}
\bar{I}_{CS}\to \bar{I}_{CS} +\frac{1}{8\pi^2} 2\pi n [\Phi (B)+\Phi(\hat B)]=
\bar{I}_{CS} +\frac{1}{8\pi^2} 2\pi n (2\pi k+ 2\pi k)= \bar{I}_{CS}+ n k.
\end{equation}
The variation is double what would be naively expected from (\ref{ICS}),
where the background contribution $\phi (\hat{B} )$ is absent
(see also \cite{Poly}). A related physical issue involves the 
requirement that the coefficient $\mu$ in $\mu I_{CS}$, viewed as a quantum 
action, be quantized. The usual argument (using $\mu I_{CS}$) is that its
 phase exponential  (the relevant quantum path integral object) must also
be (large-)invariant, so that $\mu I_{CS}$ must vary by $2\pi m, \ m\in
Z\!\!\!Z$, requiring $\mu/2\pi$ to be even. Instead, (4,5)  
imply that the parameter $\mu/2\pi$ is 
{\it any} integer \cite{Poly}. This choice leads
to a manifestly invariant complete set of states with all possible (of 
course integer) fluxes.\footnote{It has also been argued that 
consistency is preserved with a less stringent ( but still based on 
(\ref{pizza})) quantization:  specifically,
if $\mu/2\pi$ is merely rational, (only) states with the corresponding flux
values are allowed 
\cite{Poly2,00}; states
with vanishing flux are compatible with any value of $\mu$ \cite{Poly2}.  
Here, the Hilbert spaces are required to carry projective 
representations of the large gauge group. }

Let us next compare this definition of $\bar{I}_{CS}$ with the direct way of computing
the integral $\displaystyle{\int F\wedge F}$. Here we must specify the
embedding space; for simplicity, we take it to be ${\cal M}_4=D^2\times S^2$. 
[The apparent ambiguity in choice of embedding as well as of 
connection $A$, but keeping the desired boundary $\partial M_4$ and the
desired values of $A$ on it, does not affect $\bar{I}_{CS}$, the differences
being at most integer-valued.  For example, different embeddings
differ by the Chern class of the manifold obtained by gluing them 
together \cite{DW}.]
The angles ($\theta$, $\phi$) span the $2-$sphere $S^2$ while $(r,t)$ are the
polar (radial and angular) coordinates  parameterizing the disc $D^2$. 
Our desired $3-$space is the boundary $S^1\times S^2$. A 
nontrivial gauge connection $A$ on this manifold is then realized by 
requiring its  (integer) flux $\displaystyle{\int_{S^2} F}$
through $S^2$ to be  nonvanishing,
entailing  nontrivial transition functions
between the different charts covering the sphere. At the simplest 
level, we use two charts, splitting  $S^2$ into  two cups $H_{\pm}$
intersecting at some latitude $\theta=\theta_0$  and assign $U(1)$
connection 1-forms to each,
\begin{equation}
\label{coppola}
A= A_\pm + d \psi_\pm \;\;{\rm on}\;\; H_\pm.
\end{equation}
The transition function corresponding to nontrivial
flux corresponds to
 $\exp(i\psi_+)=\exp(i k \phi)\exp(\psi_-)$, which implies $A_+-
A_-= k d\phi$, ({\it i.e.} $\Phi=2\pi k$.). Regularity also requires   all 
fields to  be periodic in the angular variable $t$, with period $\beta$.
We are ready now to perform the integration, for which we revert to index
notation.
\begin{eqnarray}
&&4 \int_{D^2\times S^2}F\wedge F=\int_{D^2\times S^2}
dr dt d\theta d\phi~\epsilon^{\lambda\mu\nu\rho}F_{\lambda\mu} F_{\nu\rho}=
2 \int_{D^2\times S^2}
dr dt d\theta d\phi~\epsilon^{\lambda\mu\nu\rho}\partial_{\lambda}
\left(A_{\mu} F_{\nu\rho}\right)=\nonumber\\
&&2 \int_{S^1\times S^2}\!\!\!\!
dt d\theta d\phi\int^1_0 dr~\epsilon^{r\mu\nu\rho}\partial_r
\left(A_{\mu} F_{\nu\rho}\right)+2{\int_{S^2}
d\theta d\phi\int^1_0\!\!\!\! dr\int^\beta_0\!\!\! dt\epsilon^{t\mu\nu\rho}\partial_t
\left(A_{\mu} F_{\nu\rho}\right)}
\nonumber\\
&&+2 {\int_{D^2}dt dr\int_{S^2}
d\theta d\phi~\epsilon^{i\mu\nu\rho}\partial_i
\left(A_{\mu} F_{\nu\rho}\right)},
\end{eqnarray}
where $i\equiv (\theta,\phi)$. The first integral in the final equality
produces ( upon restoring the required normalization) the naive CS action
of (\ref{ICS}), 
\begin{equation}
I_{CS}\equiv {1\over 16\pi^2} \int_{S^1\times S^2} dt d\theta d\phi~\epsilon^{r\mu\nu\rho}
 A_{\mu} F_{\nu\rho} (r=1);
\end{equation}
the contribution at $r=0$ vanishes since $A$ is a regular connection on the
disc. The second integral is zero  since the integrand is periodic in 
$t$. However, the last term 
\begin{eqnarray}
&&2\int_{D^2}dt dr\int_{S^2}
d\theta d\phi\epsilon^{i\mu\nu\rho}\partial_i
\left(A_{\mu} F_{\nu\rho}\right)=2\int_{D^2}dt dr\int_{S^2}
d\theta d\phi\epsilon^{i r\nu\rho}\partial_i
\left(A_{r} F_{\nu\rho}\right)+\nonumber\\
&&2\int_{D^2}dt dr\int_{S^2}
d\theta d\phi\epsilon^{i t\nu\rho}\partial_i
\left(A_{t} F_{\nu\rho}\right)+\
2\int_{D^2}dt dr\int_{S^2} d\theta d\phi~\epsilon^{i j\nu\rho}\partial_i
\left(A_{j} F_{\nu\rho}\right)=\nonumber\\&& 
4\int_{D^2}dt dr\int_{S^2} d\theta d\phi\epsilon^{i j}\partial_i
\left(A_{j} F_{r t}\right)\equiv \Delta,\ \ \ \ \epsilon^{ij}\equiv
\epsilon^{rtij},  
\end{eqnarray}
requires a more careful analysis. The 
first two integrals  in the second equality can be dropped, since $A_r$ and 
$A_t$ define two  regular scalar function on the $2-$sphere. The surviving 
term  carries 
all the non-trivial information. In fact, with our above
choice of patches, we have  
\begin{equation}
\label{Luana}
\Delta=4\int_{D^2}dt dr\left [\int_{H_+}
d\theta d\phi~\epsilon^{i j}\partial_i
\left(A_{j} F_{r t}\right)+\int_{H_-}
d\theta d\phi~\epsilon^{i j}\partial_i
\left(A_{j} F_{r t}\right)\right].
\end{equation}
Using the  Poincar\'e lemma in each cup then yields
\begin{equation}
\label{poppa}
\Delta =4\int_{D^2}\!dt dr\int^{2\pi}_0\!\!
d\phi~\left[\left( A^{+}_{\phi} -A^{-}_\phi  \right)
 F_{r t} \right](\theta=\theta_0)=
4 k\int^{2\pi}_0 d\phi\int^\beta_0 dt 
 A_{t}(\theta=\theta_0, r=1),
\end{equation}
where we have used  $A^{+}_{\phi} -A^{-}_\phi=k$, as required by
(\ref{coppola}). The final result for $\bar I_{CS}$ in this procedure
thus  reads
\begin{equation}
\label{pippo}
\bar I_{CS}=\frac{1}{16\pi^2} \int_{S^1\times S^2} dt d\theta d\phi~
\epsilon^{r\mu\nu\rho} A_{\mu} F_{\nu\rho}+
\frac{k}{8\pi^2}\int^{2\pi}_0 d\phi\int^\beta_0 dt 
 A_{t}(\theta=\theta_0)\equiv I_{CS}+{\Delta\over 32\pi^2}.
\end{equation}
We have dropped the $r=1$ argument because henceforth all
fields in (\ref{pippo}) are the $4-$dimensional ones computed on the 
$r=1$ boundary, that is, the $3-$dimensional ones. 
The above route is a realization
of the prescription of \cite{DW} as well as of the procedures of \cite{Orlando}.
Several comments about 
(\ref{pippo}) are in order: (a) it is ``small'' gauge 
invariant: in fact the new contribution depends only on the integral of
$A_t$ over $S^1$ and this quantity, like the naive CS, is small (but not 
large) invariant\footnote{That the final result is not large 
invariant even though the original 4D integral is manifestly unchanged by
all gauge transformations, is traceable to the fact that three-dimensional
fields   
differing  by a large  transformation are not gauge equivalent
as components of four-dimensional 
fields. In our case one  need merely notice that a $3D$ large transformation 
affecting the integral of $A_t$ over $S^1$ must alter the flux of the $4D$ 
field through the  disc. Recall that under $U^{\rm large}_n=\exp(2\pi n 
t/\beta)$,   $A_t\to A_t+2\pi n/ \beta$.
In  $4D$ language, this corresponds to sending
$A_t\to A_t +2\pi r n/ \beta,
{\bf A}\to {\bf A}$, which is not a gauge transformation
($\displaystyle{\int dr dt \Delta F_{rt}=2\pi n}$).}. 
(b) As advertised previously, the final result is fundamentally
dependent on the patches or more precisely on the specific intersection
between different charts. (c) Finally, although (\ref{pippo}) seems quite
different from (\ref{ICS2}),
the two  are actually the same ( and of course (\ref{pippo}) varies
exactly as in (\ref{pizza})). The equivalence can be easily shown 
by  an appropriate choice of the reference connection in (\ref{ICS2}). Take,
for example,  
$\hat A$ to be any four  dimensional connection that reduces on $\partial
{\cal M}_4$
to  $(0, {\bf\hat{A}})$  where ${\bf \hat{A}}$ is the usual instanton
of topological charge $k$ on $S^2$; then 
(\ref{ICS2}) can be shown to reduce to (\ref{pippo}). (d) This also  shows 
that, in (\ref{ICS2}), the $4D$ dependent parts of ${\bf \hat{A}}$
cancel between the two terms there;
its residual lack of invariance under large transformations is purely 
$3-$dimensional. 

 We have just noticed in the descent from $D=4$ that the correction $\Delta$ required by the naive 
$I_{CS}$ is an integral over the intersection of the patches
of the transition function modulated by $A_t$.
This  correction was derived in \cite{Poly} entirely within $D=3$
(and also related there to the above descent method)
using the machinery presented
in \cite{Orlando}. To accomplish this ``intrinsic'' process,
the cohomological aspects is carried here by the various transition functions 
of the (generally complicated) overlaps. The extra contributions beyond the 
sum over the patches $\alpha$ of $\displaystyle{\int A_\alpha\wedge F}$ in
\begin{equation}
\int A\wedge F=\sum_\alpha \int A_\alpha\wedge F +\sum_\alpha T_\alpha
\end{equation}
stand for  the various  transition region
overlap terms required cohomologically to give the improved  $\bar I_{CS}$. They
 in turn are specified by the flux $\Phi$. From this
redefinition it is then  also possible to read the desired variation of 
$\bar I_{CS}$ on a large transformation.

All the above  routes for  defining a correct CS action
rely heavily on cohomological machinery. We now relate
them to the  heuristic, ``physical'' approach \cite{US} that
recasts the naive $I_{CS}$ of (\ref{ICS}) into a ``maximally'' gauge 
invariant, discarding any  ``ill-defined'' contribution
in the process (specifically in the integrations by part).
To simplify our analysis, we confine ourselves again to the case of 
$S^1\times S^2$. It can be shown  that there is a gauge reachable by
small transformations  $U=\exp i\Omega$, 
$\Omega=\tilde{A}_{0}$ (leaving $I_{CS}$ invariant) in which, starting
from abitrary $A_\mu$, the new $A_\mu$ become
\begin{mathletters}
\begin{eqnarray}
A^{U}_0(t,{\bf x})&=&\frac{1}{\beta}
\int^\beta_0 dt^\prime A_0(t^\prime,{\bf x})\equiv{\cal A}_0 ({\bf x}),\\ 
{\bf A}^{U}(t,{\bf x})&=&
{\bf A}(0,{\bf x})-{\tilde {\bf E}} (t,{\bf x}), \ \
\tilde {\bf E}\equiv-\left(\int^t_0  dt^\prime -
\frac{t}{\beta}\int^\beta_0  dt^\prime\right) {\bf E}(t^\prime,{\bf x})
\end{eqnarray}
\end{mathletters}
In terms of these variables, the naive $I_{CS}$ has the form
\begin{eqnarray}
\label{cip44}
I_{CS}&=&2\int^\beta_0 d t \int d^2 x \left[ {\cal A}_0({\bf x}) B(t,{\bf x})+ 
\epsilon^{ij}({\tilde E}_i(t,{\bf x}) 
+ A_i(0,{\bf x})) E_j (t,{\bf x})\right ]=\nonumber\\
&=&2\int^\beta_0 d t \int d^2 x \left( {\cal A}_0({\bf x}) B(t,{\bf x})+ 
\epsilon^{ij}{\tilde E}_i(t,{\bf x}) E_j(t,{\bf x})
+ \epsilon^{ij}A_i(0,{\bf x}) \partial_j {\cal A}_0({\bf x})\right )=\nonumber\\
&=&2\int^\beta_0 d t \int d^2x \left [ {\cal A}_0({\bf x}) (B(t,{\bf x})
+B(0,{\bf x}))+ \epsilon^{ij}{\tilde E}_i(t,{\bf x}) E_j(t,{\bf x}) \right],
\end{eqnarray}
where, in the last term of the second equality, we have used ${\bf E}(t,{\bf x})=
\nabla{\cal A}_0({\bf x})-\partial_0 {\bf A}(t,{\bf x})$ and then dropped  
$\partial_0 {\bf A}(t,{\bf x})$ by periodicity. In the last equality,
we have omitted the boundary term $K\equiv\displaystyle{\int d^3x \partial_j
\epsilon^{ij}({\cal A}_0({\bf x}) A_i(0,{\bf x}))}$ coming from the integration
by parts, which is patch-dependent. Surprisingly, the  final truncated
expression (\ref{cip44}) is the correct answer. A quick way of checking this is to
choose, 
in (\ref{ICS2}), any background $\hat A$ that  reduces to ${\bf A}(0,{\bf x})$ on the 
boundary.  In other words the heuristic approach implicitly promotes
${\bf A}(0,{\bf x})$ to be our reference connection $\hat A$. However, this simple
``derivation'' really involves an 
unjustified choice: the amount of ``bad term'' that we 
have to throw away is not uniquely defined. Before integrating 
by parts, the last term  in the second equality of (\ref{cip44}) involves
$\partial_j {\cal A}_0({\bf x})$ and so does not depend on 
the constant part $a$ of ${\cal A}_0$, while this dependence is restored
( by hand) 
after the integration. This mismatch obviously arises as a consequence of 
having dropped the specific boundary term $K$. However, since a part proportional to 
$a$ is well-defined irrespective of ${\bf A}$'s jumps, the amount of $a$ that 
goes into  the action or into  the boundary contribution cannot be decided 
merely from requiring a well-defined final result.

Our discussion so far has been enterely abelian. Our final derivation will 
take advantage of a simplification available in the nonabelian context of 
simply connected group such as $SU(N)$, where all $D=3$ bundles are trivial. 
This implies that there are always gauges in which the connection has no 
jumps\footnote{A non-abelian configuration 
with ``abelian'' characteristics 
that lead to apparent definition difficulties for $I_{CS}^{NA}$ is proposed 
in \cite{Roman}} and therefore the standard formula
\begin{equation}
\label{standa}
I^{NA}_{CS}=\frac{1}{16\pi^2}\int\,{\rm Tr} [A\wedge d A+\frac{i}{3} 
A\wedge A\wedge A]
\end{equation}
is valid without improvement. this fact is easy to understand in our 
$S^1\times S^2$ context because the structure of the transition function 
between the caps on $S^2$ is necessarely trivial, $\Pi_1(SU(N))=0$. Hence 
there are always sections where $A$ has been trivialized (no jumps) and 
(\ref{standa}) is applicable. Let us therefore embed our $A$ of $U(1)$ 
in $SU(2)$, by defining the $SU(2)$-valued form $A\sigma_3$. To remove the 
discontinuity in $A_{\phi}$, we have to introduce the - necessarily 
nonabelian- gauging $U$, with as usual, $A^U=U^{-1}AU-iU^{-1}dU$. 
For our model, we take
\begin{equation}
\label{Moana}
U_{+}(\theta,\phi)=\sin f(\theta) \cos n\phi I\!\!\!I + i \sin f(\theta) 
\sin n\phi \sigma_3 +i \cos f(\theta) \sigma_2\ \ \ \ \  U_{-}=I\!\!\!I
\end{equation}
where $\pm$ refers to the two caps on $S^2$ and $f(\theta)$ is monotonic regular
function so that: $f(\pi/2)=\pi/2$ and $f(0)=f^{\prime}(0)=f^\prime(\pi/2)=0$. 
At this point, $A^U$ is no longer abelian of course and we must keep both terms in (\ref{standa}). The standard 
gauge transformation rule for $I_{CS}^{NA}$ is \cite{DJT}
\begin{eqnarray}
I_{CS}^{NA}[A^U]&=&I[A]+\frac{i}{16\pi^2}\int\,{\rm Tr} 
[\,d(A\wedge dU U^{-1})]+
w(U),\nonumber\\
w(U)&=&\frac{1}{24\pi^2}\int \,{\rm Tr} [U^{-1}dUU^{-1}dUU^{-1}dU].
\label{cocca}
\end{eqnarray}
For us, $I[A]=I_{CS}[A]$ is just the naive abelian form (\ref{ICS}), so the 
complete, well-defined, result is to take $\bar{I}_{CS}=I_{CS}^{NA}$. Next, 
we observe that the winding number contribution $w(U)$ vanishes since 
it involves an explicit $\partial_t$ and the $U$ of (\ref{Moana}) is 
time-independent. The equality of the remaining term in (\ref{cocca}) with 
$\Delta$ of (\ref{Luana}) is easily verified by direct computation. 
The difference between our "secretly abelian" and truly nonabelian
configurations is also manifested by the fact that for us $w(U)$ vanished,
whereas there it is its nonvanishing that requires $\mu$-quantization 
\cite{DJT}, rather than the effect of $I_{CS} +\Delta$ noted earlier. 
In term of the language of our initial analysis, the role of $U$ is 
essentially that of the background $\hat{A}$.

To summarize, we have been comparing, from different points of view, a set of
topological and cohomological issues encountered in the analysis of the 
abelian $CS$ term at finite temperature. We have shown  how 
transgression  naturally allows us to define $\bar I_{CS}$ on nontrivial 
bundles, which are unavoidable in interesting (non-vanishing flux)
configurations, and to  easily reproduce its behavior under large gauge 
changes; we have compared this 
approach with previous ones given  in the literature and also shown it 
to underline a simple but correct heuristic definition. Finally, we have 
availed ourselves of the cohomological properties of simply connected 
groups by embedding $U(1)$ in $SU(2)$; the resulting $\bar{I}_{CS}$ immediately 
produced the desidered improvement.

\bigskip
This work is supported by NSF grant PHY93-15811, 
in part by funds provided by the U.S. D.O.E.
under cooperative agreement \#DE-FC02-94ER40818
and by INFN, Frascati, Italy.

\end{document}